\documentstyle[11pt,paspconf,epsf]{article}

\begin{document}

\title{Wide Field Observations of the Ursa Minor dSph galaxy}

\author{D. Mart\'\i nez-Delgado \& A. Aparicio}
\affil{Instituto de Astrof\'\i sica de Canarias, 38200 La Laguna, Tenerife, Spain}








\section { Introduction}

The discovery that the Saggitarius dSph is dissolving into
the Galactic halo (Ibata, Gilmore \& Irwin 1994; Mateo {\it et al} 1998)
has strengthened the idea that current accretion events could play an
important role in the history of the Milky Way formation. Nevertheless,
the argument that other dSph satellites can be tidally disrupted by
the Milky Way is open to debate. It is thus very important to
investigate the structure and history of these dSphs, including stellar
populations gradients, traces of recent star formation and the presence
of tidal debris in their outer regions. This kind
of studies using CMDs is very challenging due
to the large angular sizes and low surface density of dSphs and require
using wide field cameras and a careful analysis of the
foreground contamination.

Ursa Minor (UMi) is one of the closest satellites of the Milky Way (d=69
kpc). It is possibly a disrupted dSph interacting with the external
Galactic halo. This makes its study quite necessary in the forementioned
context. In this paper we
present preliminary results of a wide field photometry survey of
UMi and discuss the presence of an intermediate-age population and
tidal tails in it.

\section{Observations}

UMi dSph was observed in $B$ and $R$ with the Wide Field Camera (WFC) at the prime focus of the INT (2.5m) at the Roque de los Muchachos Observatory (Canary Islands). The WFC holds four 4096 $\times$ 2048 pixels EEV CCDs with a pixel size of 0.33 $\arcsec$, covering a total area of 0.27 square degrees for field. Figure 1 shows the fields selected for the first observing run of this project.

%

\section{The color-magnitude diagram of Ursa Minor dSph}

Figure 2 shows the [$B-R$, $"V"$] CMDs for the central body of UMi dSph
(field A of Fig. 1). These diagrams show a prominent narrow RGB
suggesting a small metallicity range in the stellar population. In addition,
a well populated, blue extended horizontal branch and several ($\sim$
150) stars in the RR Lyrae gap region are observed. The main turn-off
corresponds to an age of $\sim 15$ Gyr for $Z=0.0004$, based on the
Padua's stellar evolutionary models (Bertelli et al. 1994, and
references therein). 
The few stars above the subgiant track ($0.6 \leq (B-R) \leq 1.0$; $
20.5 <"V"< 22.5$) are probably binary stars although they could be the
trace of a not so old population. 

Perhaps the most streaking feature is the well populated {\it blue
plume}. A simple explanation is that it would be made of blue
stragglers. However, normalized to the RGB, it contains three times more
stars than low central density Galactic globular clusters, like NGC 5053
(Nemec \& Cohen 1989). This perhaps indicates that the blue-plume is the
trace of an intermediate-age population. Nevertheless, photometry alone
is not enough to distinguish blue stragglers from normal main sequence
stars. A definitive answer will requires spectroscopic data.

\begin{figure}
\plotfiddle{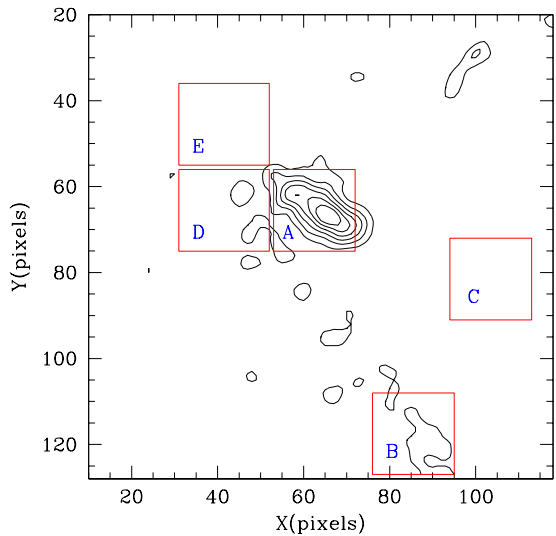}{5cm}{0}{40}{40}{-150}{-80}
\caption{Location of the WFC fields of our study of UMi. North is to the
top and East to the left on the isopleth map by Irwin \& Hatzidimitriou
(1995). The scale is 1.65 $\arcmin$/pixel.} \label{fig-1}
\end{figure}

\section{ An intermediate-age population in UMi?}

We have done a preliminary estimate of the star formation history for
the central body of UMi using the CMD of chip 4 (Figure 2) and assuming
that the blue stars present in the CMDs are not blue-stragglers but an
intermediate-age population. The result is shown in Figure 3 (left
panel). A high SFR is required in an early epoch to account for the
amount of HB and RGB stars. After $\sim 12$ Gyr ago, a low star
formation activity extended until a recent epoch, is enough to produce
the observed blue-plume. Interestingly, UMi would have maintain this
intermediate-age SFR if it  somehow would have manage to convert into
new stars only a 5$\%$ of the material returned to the interstellar
medium by older stellar generations. 

For illustrative purposes, right panel of Figure 3 shows the corresponding synthetic CMD. For simplicity, a constant metallicity ($Z=0.0004$) has been
used and no observational effects have been simulated.

\begin{figure}
\plotfiddle{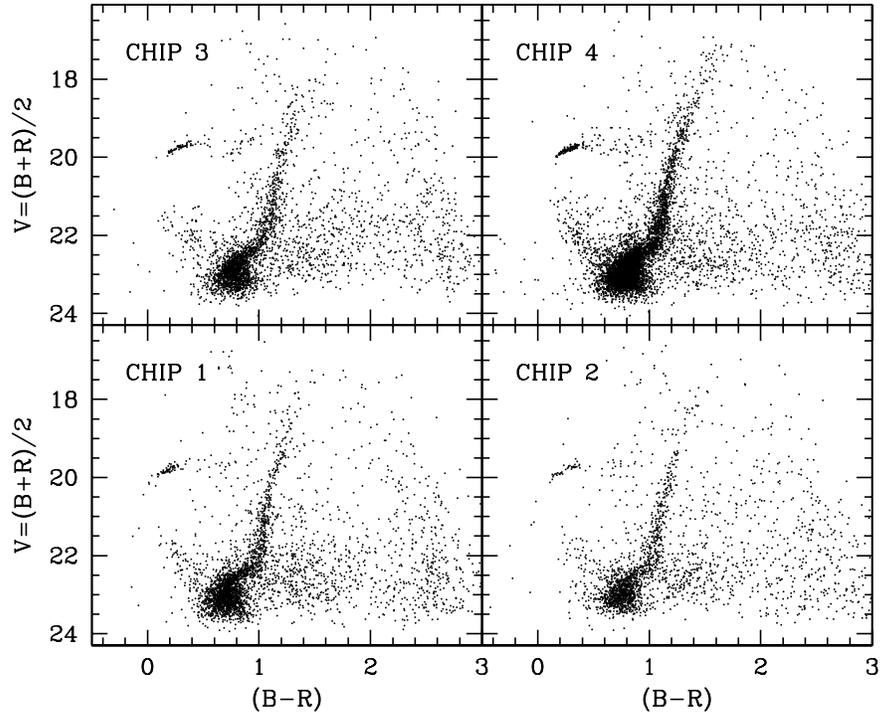}{8cm}{0}{60}{60}{-200}{-100}
\caption{CMDs for the central body of UMi (field A). Each diagram
corresponds to one of the four chips of the INT WFC. Chip 4 includes the
nucleus of the galaxy.} \label{fig-1}
\end{figure}

\begin{figure}
\plottwo{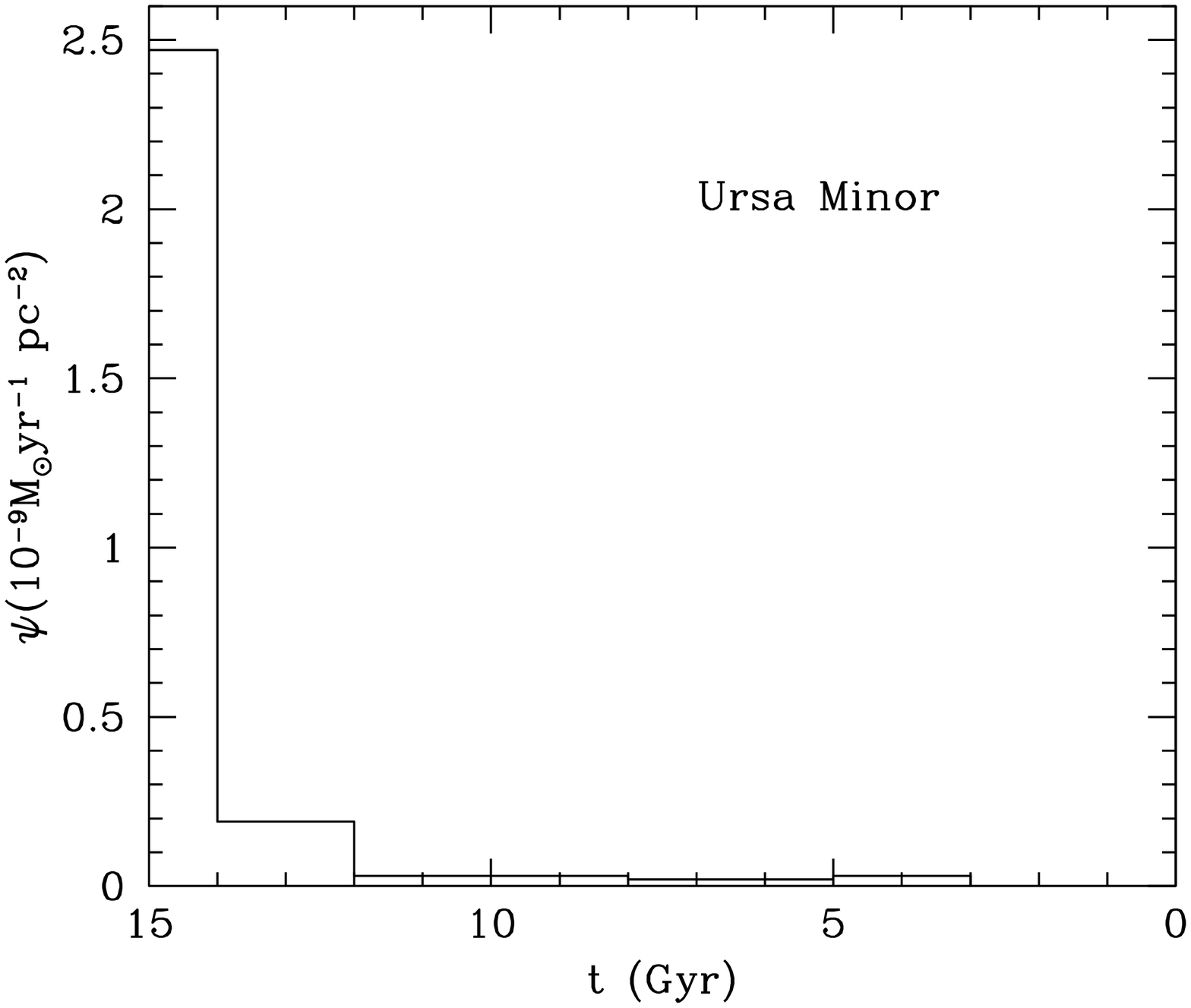}{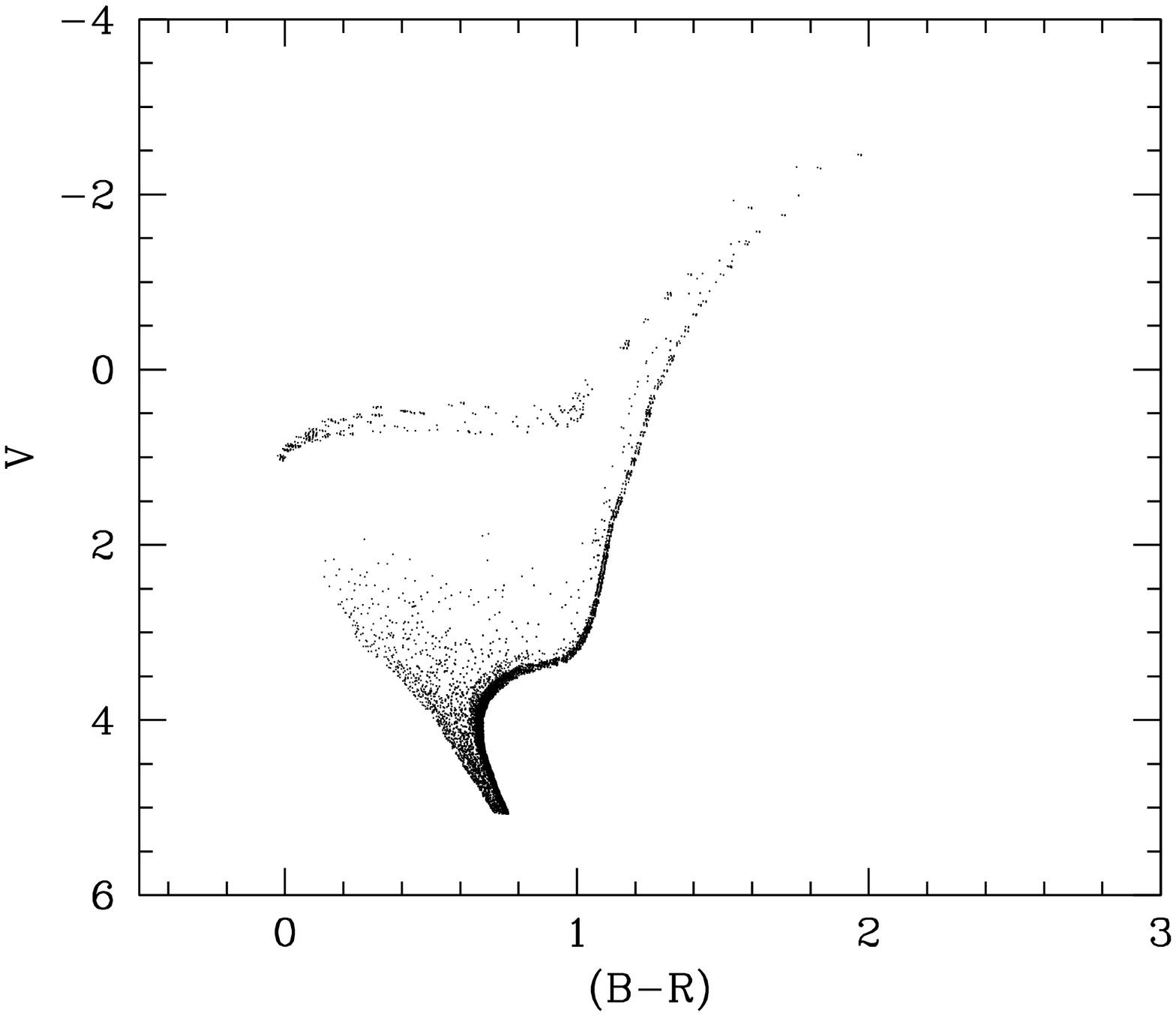}
\caption{The SFR computed for the central body of UMi (left) and the
corresponding synthetic CMD (right).} \label{fig-1}
\end{figure}

\section{Searching for tidal tails}

As a part of our project, we are searching for tidal tails in the
periphery of UMi. A first test is the detection of relevant structures in
the CMDs of these regions, such as RGB or HB traces. Figure 4 shows the
CMDs of field C (shown in Fig. 1), situated
over the major semiaxis of UMi and  beyond its tidal radius in the direction of the Magellanic Stream. No traces of the
galaxy are found in this field. A similar result is found for the field
B (see Fig. 1).	The lack of traces beyond the tidal radius may indicate
the absence of extra-tidal stellar debris and suggests that UMi would not
be in an advanced state of tidal disruption.

\begin{figure}
\plotfiddle{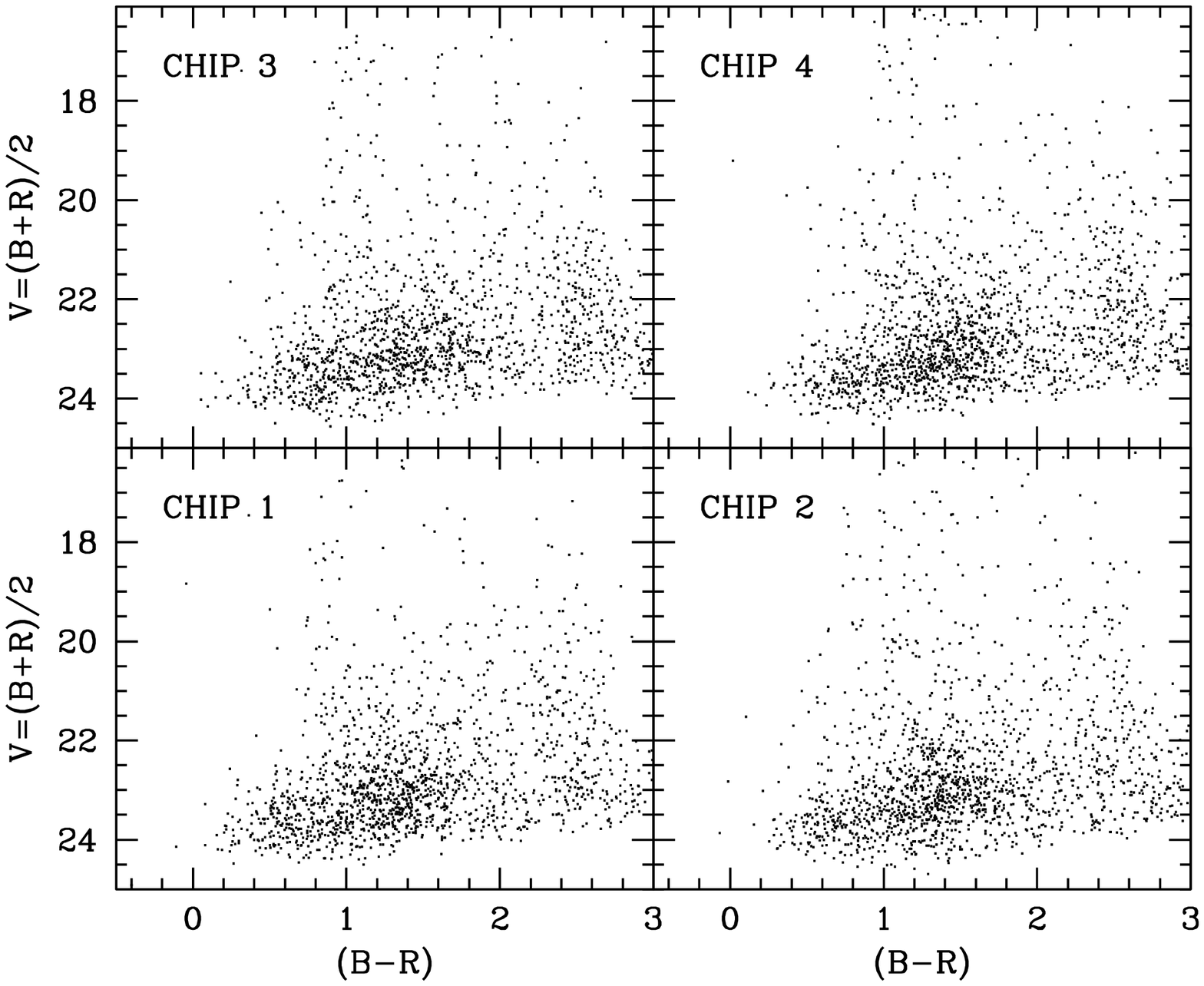}{8cm}{0}{60}{60}{-200}{-100}
\caption{CMDs for the field C (see Fig. 1) situated in the outer
regions of UMi, beyond its tidal radii. No trace of the galaxy is found.} \label{fig-1}
\end{figure}

\end{document}